\documentstyle[12pt,epsfig]{article}
\font\tenrm=cmr10  
\def\aprle{\buildrel < \over {_{\sim}}} 
\def\aprge{\buildrel > \over {_{\sim}}}

\begin{document}     
\title {On Exotic Solutions of 
the Atmospheric Neutrino Problem.}
\author{Paolo Lipari and Maurizio Lusignoli \\
Dipartimento di Fisica, Universit\`a di Roma ``la Sapienza",\\
and I.N.F.N., Sezione di Roma, P. A. Moro 2,\\ I-00185 Roma, Italy}

\date{~~~~}      

\maketitle
\begin{abstract}  
The measurements of the fluxes of atmospheric  neutrinos
give  evidence for the  disappearance of  muon  neutrinos.
The  determination of the   dependence of the disappearance 
probability  on the  neutrino  energy and  trajectory allows
in  principle to   establish  unambiguosly the  existence of 
neutrino oscillations.
Alternative  mechanisms  for the  disappearance   of the  neutrinos
have been  proposed,  but  do  not  provide  a  viable   description  of
the  data, if one includes both events where the neutrinos
interact in the detector  and  $\nu$--induced upward going  muons.
The   proposed  mechanisms   differ in the   energy dependence of the
disappearance probability   and the   upward going  muon  data
that are produced by  high energy neutrinos  give  a crucial
constraint.
\end{abstract}   

\section {Introduction} 
The measurements of the  fluxes of  atmospheric  neutrinos
by the  Super--Kamiokande (SK)  experiment  
\cite{SK-evidence,SK-subGeV,SK-multiGeV,SK-upmu,SK-Takayama} show  evidence
for the disappearance of muon  (anti)--neutrinos.
The same indication comes from the older data of the Kamiokande \cite{oldK}
and IMB \cite{IMB} experiments and the recent ones of Soudan--2
\cite{Soudan}. Also the results recently presented by the MACRO collaboration 
\cite{MACRO-upmu,MACRO-Takayama}
indicate a suppression of the muon (anti)--neutrino flux.

The simplest explanation  of the  data  is the  existence of
$\nu_\mu \leftrightarrow \nu_\tau$ oscillations \cite{SK-evidence}.
In the framework of flavor oscillations one  should 
consider  the more  general case of three flavors
\cite{3-flavors}
(with the Chooz  experiment  \cite{Chooz} giving important
contraints  to the electron neutrino  transitions),
and could  also  envisage more complex  scenarios  involving 
sterile states \cite{sterile}.   We will not pursue  these  
possibilities here, and we will  adopt instead the  simplest
scenario  of  two--flavors oscillations as a prototype model that, as
we will see, is able to describe successfully the experimental data.

We will instead investigate  if  other  forms  of  `new physics' 
beyond the standard model, different from standard flavor oscillations,  
can also provide  a satisfactory   description 
of the  existing data.
Indeed several other physical mechanisms  have  been  proposed in
the literature as viable explanations of  the atmospheric  neutrino data.   
In this  work  we will consider three of these  models:
neutrino decay \cite{nu-decay},  
flavor  changing neutral  currents (FCNC) \cite{Broj,FCNC},
and violations of the equivalence  principle
\cite {Gasperini,HL} or, equivalently, of Lorentz
invariance  \cite{GHKLP}.
All these model  have the common feature of `disappearing'  muon neutrinos,
however the   probability depends in different  ways  on the 
neutrino energy and  path. To discriminate between  these models
a  detailed study of the disappearance  probability  $P$  and of 
its  functional form is needed. 

In this  work in contrast with  previous  analyses
we will argue that the present data allow  to exclude 
the three  `exotic'  models,
at least in their simplest form,
as explanations of the atmospheric  neutrino problem.
This is mainly due to the difficulty that
these models have to fit at the same time the SK data for leptons 
generated inside the detector (sub-- and multi--GeV) and for up--going muons
generated in the rock below it.

\section {Data}
In fig. 1, 2 and 3  we  show (as  data points  with statistical error bars)
the ratios between the SK data and their Montecarlo  predictions 
calculated  in the  absence of oscillations or other
form of `new physics' beyond the standard model.
In fig.~1  we show  the data for the  $e$--like contained  events,
in fig.~2  for $\mu$--like events produced in the detector,
and  in fig.~3  for upward-going muon events,
as a function of zenith angle  of the detected  lepton.
In each figure we  include four lines: the dotted line 
(a constant of level unity)  
corresponds to exact agreement between  data  and
no--oscillation  Montecarlo, including the absolute normalization.
The dot--dashed  lines correspond
to the assumption that there is  no deformation  in the shape
of the zenith angle  distributions, but that one is allowed to change
the normalization of each data sample  independently.
The values obtained are:
1.16 for  $e$--like  sub-GeV,
1.21 for  $e$--like  multi-GeV,
0.72 for  $\mu$--like  sub-GeV,
0.74 for  $\mu$--like  multi-GeV,
0.56 for  stopping upward-going muons,
and 0.92 for  passing upward-going muons.
For two sets  of data (sub-GeV and  multi-GeV $\mu$--like events)
the constant shape fits  give very poor descriptions 
($\chi^2 = 26$ for  the  sub-GeV   and 33 for the multi-GeV for 4 d.o.f).
Also  the zenith angle shape of the passing upward-going muons
is not well  fitted  by  the no--oscillation Montecarlo
($\chi^2 = 17$ for  9 degrees of  freedom).
The electron  data    do not  show  clear evidence of  deformations,
although the    constant shape   fit  for the sub-GeV  events 
($\chi^2 = 9.7$ for  4 d.o.f.) is  rather  poor.

The normalizations of the different data  sets  are of course
strongly correlated, and therefore 
it is not reasonable to let them  vary independently.
The other extreme option, that we will adopt
in this  work  for  simplicity, is to use one and   
the same parameter to fix the normalization of the
six data samples. The result for constant shapes (i.e. assuming no
`new physics' beyond the standard model) 
is represented by the dashed lines in fig. 1, 2 and 3
corresponding to a value 0.84 and a very poor $\chi^2=280$ for 34 d.o.f.).

The full lines in the figures correspond to our best fit assuming
$\nu_\mu \leftrightarrow \nu_\tau$ oscillations with maximal mixing.
We define  the $\chi^2$ as follows:
\begin {equation}
\chi^2 = \sum_j \left [ { N_j - \alpha \;
N_j^{th} ~(N_{j,MC}^{SK}/N_{j,0}^{th}) \over \sigma_j } \right]^2
\label{eq:chi2}
\end{equation}
In (\ref{eq:chi2}) the summation runs over all data bins, 
$N_j$ is the  SK  result  for the $j$--th bin,
 $\sigma_j$ its  statistical error, 
$N_j^{th}$ our prediction,
$N_{j,0}^{th}$ our prediction  in the absence of oscillations, 
$N_{j,MC}^{SK}$  the no--oscillation  prediction of 
Super--Kamiokande, and $\alpha$  allows  for variations in 
the absolute   normalization of the prediction.
We have rescaled  our  prediction to the SK   Montecarlo because we do 
not have a  sufficiently  detailed  knowledge   of the
detector    response (e.g.  number of detected  rings) and efficiency.
For the same same input  neutrino spectra
the difference between  our no--oscillation calculation
(see \cite{ADLS}  for  a  description)  
and the SK  Montecarlo  result is  approximately 10\%, 

For our best fit 
the values of the relevant parameters are $\alpha = 1.15 $ and 
$\Delta m^2 = 3.2\;10^{-3}$~eV$^2$. The $\chi^2$ is 33.3 for 33 d.o.f. 

Our  definition of the $\chi^2$ is  somewhat  simplistic.
We  do not take into account the contribution of
systematic errors,  either in the data
or in the theory.  The assumption of a common   $\alpha$ for
$e$--like  and $\mu$-like  events   corresponding to different
energy regions is  certainly too strict. It is therefore remarkable
that this fit is so good, and essentially in agreement (same normalization
and very near $\Delta m^2$ value) with the much more elaborate fit
in \cite{SK-evidence}.

In the rest of this paper we will consider other, `exotic' models and
we will find that they are not able to provide a satisfactory fit to
the same data.

\section{Models}
We briefly recall the essential points of the models we are discussing.

For the usual, two--neutrino flavor oscillations the
`disappearance  probability' $P$     is given by:
\begin{equation}
P = P^{osc}_{\nu_\mu \to \nu_\tau}  = 
\sin^2 2 \theta ~\sin^2 \left [ {\Delta m^2 \over 4} \; {L \over E_\nu}
\right ],
\label{P-osc}
\end{equation}
with  the   very characteristic sinusoidal dependence on  the  ratio
$L/E_\nu$. 

In the simplest  realization of  neutrino  decay,
neglecting the possibility
of the simultaneous existence of neutrino oscillations, the   
disappearance probability is given by:
\begin{equation}
P = P^{dec} = 1 - \exp \left
[- {m_\nu \over \tau_\nu} {L\over E_\nu} \right ],
\label{P-decay}
\end{equation}
still depending on the ratio between neutrino pathlength and energy
$L / E_\nu$, but with a functional form different from (\ref{P-osc}).

If  flavor  changing  neutral  currents  contribute to the 
interaction of neutrinos with ordinary matter,  
a non  trivial  flavor  evolution will develop even for massless neutrinos
as  originally   noted  by Wolfenstein \cite{Wolfenstein}.
There are several theoretical models generically predicting 
nondiagonal neutrino interactions with matter. In 
particular such models have been proposed 
as a possible consequence of $R$-parity violating interactions 
in supersymmetric models  and  suggested as solutions  of both the
solar  \cite{Rparity,Broj}  and  atmospheric \cite{Broj,FCNC} neutrino problems.
Let us call  $V_{\alpha\beta}$ the  effective potential
that  arises    from the forward scattering   amplitude 
of a neutrino with a fermion $f$: $\nu_{\alpha} + f \to  \nu_{\beta} + f$.
In the standard model    
$V_{\mu\tau} = V_{\tau\mu} = 0$, and 
$V_{\mu\mu} = V_{\tau\tau} = \sqrt{2}\,G_F\, T_3(f_L)\, N_f$   where 
$G_F$ is the Fermi  constant, $N_f$  is the number density of the fermion $f$
and  $T_3(f_L)$ is the third component of the fermion's weak isospin.
Since the  effective  potentials  for  muon and  
tau neutrinos  are identical,  there is  no  effect on  standard oscillations.
However, if  
the scattering  amplitudes  are  different  from those predicted by 
the standard model, 
and if flavor  changing  scattering  can occur, then the 
effective  potential  acquires  non diagonal terms 
$V_{\mu\tau} = V_{\tau\mu} = \sqrt{2} G_F \epsilon N_f$,
and different diagonal elements (with 
$V_{\tau\tau} - V_{\mu\mu} = \sqrt{2} G_F \epsilon' N_f$),
and there  will be  a nontrivial flavor transition probability  even for 
massless neutrinos.
After  the crossing  of a layer  of  matter  
with a column density
\begin {equation}
 X_f = \int_0^L dL' ~N_f(L') \;,
\end{equation}
the  transition probability  is:
\begin{equation}
P = P^{FCNC}_{\nu_\mu \to \nu_\tau}  =
 \frac{4\epsilon^2}{4\epsilon^2+\epsilon'^2}\;
~\sin^2 \left [ {G_F \over \sqrt{2}}\,  X_f \,\sqrt{4\epsilon^2+\epsilon'^2} 
 \right ].
\label{P-FCNC}
\end{equation}
The probability has  again an oscillatory form, 
however in this case the role 
of  $L/E_\nu$  is taken by the column  density  $X_f$  and there
is no dependence on the neutrino  energy.

If the gravitational  coupling of neutrinos  are  flavor dependent
(implying a violation of   the equivalence  principle)
mixing  will take  place  for   neutrinos
traveling in a gravitational  field    even for  massless  neutrinos
\cite{Gasperini,HL}.
The  neutrino states with  well defined  coupling to the gravitational field
define   a  `gravitational basis'
related  to the flavor basis
by a unitary  transformation.
The effective  interaction energy matrix of neutrinos in  a 
gravitational  field can be  written in an arbitrary basis as
\begin {equation}
  H = -2 \,|\phi(r)| ~E_\nu  (1 + f)
\end{equation}
where  $E_\nu$ is the  neutrino  energy,  $\phi(r) = - |\phi(r)|$ is
the gravitational potential, and  $f$ is   a (small, traceless)  matrix
that  parametrize the  possibility of  non--standard  coupling 
of  neutrinos  to gravity  and is diagonal in the gravitational basis.

Much in the same way as in the previous cases, 
the noncoincidence of gravitational and flavor eigenstates 
determines mixing and flavor transitions. 
Considering the  simple  case of two  flavors
and  assuming  a constant  gravitational  potential $|\phi|$,
the  transition probability takes the  form
\begin{equation}
P = P^{grav}_{\nu_\mu \to \nu_\tau} =
\sin^2(2 \theta_G) \;
\sin^2[ \delta |\phi| ~E_\nu \,L ].
\label{P-equiv}
\end{equation}
where
$\theta_G$ is  the  mixing  angle  
and $\delta$  is  the difference
between the  coupling to gravity   of the  gravitational  eigenstates.
Note that in this case the argument of the oscillatory function
is proportional to the product of the neutrino energy and 
pathlength, whereas for the standard flavor oscillations it is the
ratio  of the same quantities that matters.

Equations (\ref{P-osc}),  (\ref{P-decay}),  (\ref{P-FCNC})
and (\ref{P-equiv})
are the disappearance  probabilities for the  four  mechanisms  that we will
confront with the experimental  data.

\section {Flavor Oscillations}
It is interesting to discuss    how  the usual  flavor  oscillations
can successfully reproduce the  pattern of suppression  measured 
for the different event  samples.
The  events detected in one  particular   bin are  produced 
by  neutrinos with a predictable distribution of
$E_\nu$ and pathlength $L$, and   therefore 
of $L/E_\nu$, the significant  quantity in 
flavor oscillations.
In fig.~4, in the top  panel  we show as  a function of
$L/E_\nu$ the survival probability  corresponding to  
maximal  mixing and $\Delta m^2 = 3.2 \times 10^{-3}$~eV$^2$
(our  best fit point).  Also  shown with a dashed  line 
is the survival probability for  neutrino decay  that  we will
discuss in the  next section.

In the second panel we  show the $L/E_\nu$ distributions   
of  sub--GeV  $\mu$--like  events 
in the five zenith angle  bins used  by the SK collaboration:
$\cos \theta_\mu \in [-1,-0.6]$, $[-0.6,-0.2]$
$[-0.2,+0.2]$, $[0.2,0.6]$ and $[0.6,1.0]$
(corresponding to the thick solid, thick dashed, thin  dot--dashed,
thin dashed and  thin solid  line).

In the third panel we  show the corresponding
distributions  for  multi-GeV  $\mu$--like  events 
(same  coding for the lines).

In the fourth panel  we  show   the $L/E_\nu$  distributions for
upward  going  muons that  stop in  the  detector 
in the   zenith  angle  bins:
$\cos \theta_\mu \in [-1,-0.8]$, $[-0.8,-0.6]$
$[-0.6,-0.4]$, $[-0.4,-0.2]$ and $[-0.2,0.0]$ 
with the  corresponding lines  ordered from right 
(higher  values of  $L/E_\nu$)  to left 
(lower values of  $L/E_\nu$).

In the last panel  we  show   the same distributions 
for passing  upward  going  muons
in ten  zenith angle  bins, 
$\cos \theta_\mu \in [-1,-0.9]$, $\ldots$, $[-0.1,0.0]$.

Some remarks can be useful for an understanding of the distributions
shown in  fig.~4.  For the sub-GeV events, one  can see
that  the parent neutrinos   have $L/E_\nu$ spread over a broad 
range of values.  This  is  due to the poor correlation between the 
neutrino and muon  directions
$\langle \theta_{\nu\mu} \rangle \simeq 53^\circ$.

For multi-GeV data the distributions
are much  narrower, reflecting the tighter  correlation between
the neutrino and muon directions, 
$\langle \theta_{\nu\mu} \rangle \simeq 13^\circ$. 
Note also  that the peaks
in the $L/E_\nu$   distributions
corresponding to  sub-GeV  and multi-GeV events  in the same
zenith angle  interval    are  at slightly different  points 
because of the  different   energy of the parent neutrinos.

For up--going stopping muons  the width of the distribution 
is  wider   than in the  multi-GeV case.
The correlation between the muon and  neutrino directions 
$\langle \theta_{\nu\mu} \rangle \simeq 10^\circ$  is actually better,
but the width of the distribution reflects
the wider energy  range of the neutrinos contributing  to this signal.
Passing muons    are  nearly collinear with the parent neutrinos
($\langle \theta_{\nu\mu} \rangle \simeq 2.9^\circ$),  but
the large energy range of the  neutrinos that extend over nearly
two decades ($E_\nu \simeq  10$--10$^3$~GeV) results  in  a 
wide  $L/E_\nu$ distribution.

All  curves in the lower  four  panels  of fig.~4  are  normalized  to unit
area.  In order to  obtain the  suppression  due  to oscillations
in a particular  bin, one has to  perform the integral:
\begin{equation}
  N_j^{osc}  (\sin^2 2\theta, \Delta m^2) = \int dx ~ {dN_j^0 \over dx }
~[1 - P^{osc}(x, ~\sin^2 2\theta, \Delta m^2) ]
\end{equation}

Comparing the survival probability
with the  $L/E_\nu$ distributions  it is  easy to gain 
a  qualitative  understanding of  the effects  produced.
For $\Delta m^2 \simeq 3\cdot 10^{-3}$~eV$^2$,   neutrinos
with $L/E_\nu \aprle 10^{2}$~Km/GeV have  a  survival  probability
close to unity  and do not  oscillate,   while   for neutrinos
with  $L/E_\nu \aprge 10^{3}$~Km/GeV,  averaging over the rapid oscillations,
the survival  probability becomes  one half for maximal  mixing.
We recall that  horizontal neutrinos 
travel an average pathlength of $\simeq 600$~Km.

Taking  into  account the $L/E_\nu$  distributions  of the different
set of events  one can  see  that all  zenith angle bins
of the muon sub--GeV   events  are somewhat  suppressed, because  
even vertically  downward going  muons  can be produced by
upgoing neutrinos.   

For  multi--GeV events, with the tighter  correlation between the neutrino
and muon  directions, the  two up--going  bins  are suppressed
by the `average' factor $\sim 0.5$,  
the  two down--going  bins  are left unchanged  and the  horizontal  muons
have  an intermediate  suppression.

The up--going stopping muons are always suppressed by a factor $\sim$ 1/2,
except for the bin nearest to the horizontal. 

For the up--going passing muons the larger average energy and therefore
smaller $L/E_\nu$  explains   the smaller suppression  
and its  pattern, varying from nearly unity
for the horizontal bin to a maximum of $\sim 0.65$ for the vertical one.

\section {Exotic Models}

\subsection {Neutrino Decay}

Fitting the sub-GeV and multi-GeV data of Super-Kamiokande  with the   
simplified  model of muon neutrino decay (that neglects mixing)  given
in (\ref{P-decay}),
we find a minimum in the $\chi^2$  for a value 
$\tau_\nu/m_\nu = 8900$~Km/GeV  (with $\alpha = 1.07$).
  This is  in good  agreement
with the results  of  \cite{nu-decay}.  The authors of this  reference
have as a best fit  point
$\tau_\nu/m_\nu  \simeq 12800$~Km/GeV, with a small 
mixing  angle  $\sin^2 2 \theta \simeq 0.06$.
The   curve  describing the decay probability for our  best fit
is  shown as the dashed  line in the top panel  of fig.~4.

It is simple to have a qualitative understanding  of the value
of $\tau_\nu/m_\nu$ that provides the best fit. One  needs  to suppress
by a  factor $\sim  0.5$ the up-going   multi-GeV   muons
that have $\langle L/E_\nu \rangle \simeq 10^{3.5}$~Km/GeV
(see fig.~4).

The inclusion of decay     results
 in   $\chi^2 = 71$  (for 18 d.o.f.), a very significant  improvement
over the value 234 (19 d.o.f.)   of the ``standard model", 
but still significantly  worse than  the value
$\chi^2 \simeq 25$  of the $\nu_\mu \leftrightarrow \nu_\tau$ 
flavor oscillation fit  to the same set of data.

For  a value of  $\tau_\nu/m_\nu$ of the order of what is  given by  
our fit to the sub--GeV and multi--GeV data, one  expects a much smaller 
suppression of the high  energy  passing up--going muons (as already noted in
\cite{nu-decay}).  In fact 
including also the 15 data points of  the  up-going muons  in a new fit, the
best fit point becomes $\tau_\nu/m_\nu = 10000$~Km/GeV (similar 
to the previous one), but  $\chi^2$  increases to  the  
much higher value 140.

\subsection{Violation of the equivalence principle}

Performing a fit to the sub-GeV and multi-GeV data of  
Super Kamiokande with the disappearance probability
given by (\ref{P-equiv}) and with maximal mixing ($\theta_G = \pi/4$),
we find a minimum in the $\chi^2$  for a value 
$\delta |\phi| = 4 \cdot 10^{-4}$ Km$^{-1}$GeV$^{-1}$ (with $\alpha = 1.10$).
The $\chi^2$ for this  fit is 35 for 18 d.o.f., 
still a very significant improvement over the standard model case, 
but not as good as the flavor oscillations result.
The survival probability given by our best fit 
is shown in the top panel of fig.~5. 

The reason of  the poor  $\chi^2$  can  qualitatively  be understood
looking at fig.~5.
This  figure is the equivalent of 
fig.~4,  in the sense that the four lower panels  show the  distributions
in the variable that is relevant in this  case, namely $L\cdot E_\nu$.
The  distributions in this  variable  for the sub--GeV  and multi--GeV
events   have  shapes similar  to the corresponding ones in 
$L/E_\nu$, because the width of the distributions  is 
mostly  determined  by the spread in pathlength $L$.
However  the    average value of   the $L\cdot E_\nu$ 
of the   sub--GeV  events
is lower than the corresponding one
(same zenith  angle bin) for  multi--GeV events,
the opposite of what happens in the  $L/E_\nu$ distributions, see fig.~4.
Therefore, parameters describing well multi--GeV events
will generally produce too low a suppression for sub--GeV events or
viceversa. 

It can  be argued (as the  authors of reference \cite{fly}  do)
that   taking into account systematic  uncertainties 
the model  defined  by equation (\ref{P-equiv})  provides  a  good
fit to the data, however this is  not the case if upward--going muons
are  included in the picture. This should be  evident  looking
at the lower panels  in fig.~5.  Upward--going muons are produced
by high energy neutrinos  and the frequent  oscillations  do imply
a suppression by   50\% of   passing  (and stopping) muons,
with no  deformation of the zenith angle distribution.
This is in   disagreement 
with the corresponding data: in fact, trying to fit all the data 
together we obtain similar best fit parameters,  
$\delta |\phi| = 4.5 \cdot 10^{-4}$ Km$^{-1}$GeV$^{-1}$ and
$\alpha = 1.145$, but with
a very bad $\chi^2 = 142.7$ for 32 d.o.f. (the 
contribution of passing upward--going muon data being $\sim 100$).

\subsection{Flavor Changing Neutral Currents}

In  the case of  neutrino  transitions produced  by 
flavor  changing  neutral currents, the  r\^ole
of $L/E_\nu$ is replaced  by $X$, the  column density.
This has  the fundamental consequence that 
there is no energy dependence  of the flavor  conversion.
Moreover since  air has  a density much lower that
the Earth's, the  transitions do not  develop 
during the   neutrino path in the atmosphere, and therefore down--going
neutrinos  are unaffected.
Note also  that there is not a  simple   relation   between 
the zenith angle $\theta_\nu$  and  the pathlength $L$
because of fluctuations  in the  neutrino birth position.
However, due the  air low density, the   zenith angle
$\theta_\nu$  does define the  column density $X$  
with a negligible error: the entire  down--going hemisphere 
corresponds  to   $X \simeq 0$ and to  a vanishing transition probability.

Performing, as before, a  fit to the sub-GeV and multi-GeV data of  
Super Kamiokande with the disappearance probability
given by (\ref{P-FCNC}) and assuming scattering off down quarks and
$\epsilon^\prime = 0$
(that is maximal  mixing), we 
obtain a  best  fit   value $\epsilon = 0.4$ and
$\alpha = 1.08$
corresponding to a
minimum $\chi^2 = 38$.
With increasing $\epsilon$ the   oscillations   become  more frequent,
and  essentially  all  values $\epsilon \aprge 0.4$ give  comparable fits,
since for these large values  the oscillations  
can be considered as  averaged  in the  entire up--going hemisphere.

The authors of reference \cite{FCNC},  exploring the parameter space
($\epsilon,\epsilon^\prime$)  find  two solutions:
(a): (0.98,0.02) and  (b): (0.08,0.07), that are plotted 
in the upper  panels  of fig.~6.
The first solution corresponds   to the one  that we have found,
considering the   slow  variation  of $\chi^2$ with $\epsilon$ in the
large  $\epsilon$ region. 
The $\chi^2$   found by the authors of \cite{FCNC}  is  however
better  that   what we  find,  indeed  as  good as 
in the  flavor oscillation  model.

We do  find  that  fitting  the muon data only,    without  considering the
constraint on the normalization  coming from the electron data,
the FCNC  model  gives  an excellent fit, indeed  as  good   or  better
than the flavor oscillation  model.
The  reason  why, in our fitting  procedure, 
the  FCNC model gives not as  good a fit  
originates from the fact that the  theoretical average 
value  of the suppression  for  both
sub--GeV and multi--GeV    muon events for the  best fit parameters
is $\simeq 0.75$,  corresponding to no suppression in
the down--going   hemisphere  and $\sim 0.5$ in the opposite one.
The data \cite{SK-subGeV,SK-multiGeV}
for  the double ratio  $R = (\mu/e)_{Data}/(\mu/e)_{MC}$:
$R_{sub} = 0.61 \pm 0.03 \pm 0.05$, and
$R_{multi} = 0.66 \pm 0.06 \pm 0.08$ 
indicate  a larger  average  suppression.
The allowance of a non perfect correlation 
between  the normalizations of the muons
and  electron data samples 
would certainly reduce   the $\chi^2$ value of our fit.

The  inclusion of  up--going muons  among the data 
considered,  again  results  in evidence against 
this  model.
We  recall the fact that  the passing muons are
essentially collinear  with the parent neutrinos
and that the experimental zenith angle distribution
does  not exhibit   large sharp  features  as those  predicted
for example  by solution (b)  of \cite{FCNC}  (see fig.~6).
Therefore  the relative smoothness  of the passing muon
data allows  to exclude  a large range of  values 
($\epsilon,\epsilon^\prime$)   that   correspond to
few oscillations in the up--going hemisphere 
(that is $0.04 \aprle \sqrt{4\epsilon^2+\epsilon'^2} \aprle 0.2$)
and  still large  effective  mixing
$\epsilon \aprge \epsilon^\prime$.

The solution (a)  of \cite{FCNC} cannot  be excluded  using this
consideration, because   its  frequent  oscillations   do  not 
produce  sharp  features   given the 
binning  of the experimental  data, and   give
a constant suppression $2 \epsilon^2/(4 \epsilon^2 + \epsilon'^2)$
for all  zenith angle  bins.
The model  has  no energy dependence,
and  therefore  this  average suppression  must apply  to
the up--going passing  and stopping  events, as  well as
to the up--going multi--GeV events,  that have  also
a  rather  sharp  correlation between  the neutrino and muon
directions.
This is  in   disagreement with two features  of the experimental data:
(i)  the passing muons  have  a suppression  considerably less
than  both the stopping  and up--going multi--GeV muons;
(ii)  the shape of the zenith  angle distribution of passing muons
shows  evidence for a deformation.
More quantitatively, a fit to all the data  with $\epsilon^\prime = 0$
gives the parameter values $\epsilon = 1.4$ and $\alpha = 1.12$ 
but with a total $\chi^2 = 149$ (the contribution 
to $\chi^2$ of the throughgoing muon data being 105).

\section{Summary and conclusions}

The  survival probability  $P(\nu_\mu \to \nu_\mu)$ in the case
of  two flavor $\nu_\mu \leftrightarrow \nu_\tau$ 
oscillations   has  a well  defined  dependence on the pathlength and  
energy of the neutrinos.
In order  to  establish  unambiguosly  the existence of such  oscillations
it is necessary to   study in  detail  these dependences.
In the analysis of the events interacting in the detector,
one  can study a very  wide  range of  pathlengths
($10 \aprle L \aprle 10^{4}$~Km)    but a much smaller
range of neutrino  energies   close 
to 1~GeV  (the sub--GeV and multi--GeV samples).
Therefore it is  not  easy to obtain experimental information on
the  dependence  of the  survival probability
on the neutrino energy.
In fact models    where the combination
$L/E_\nu$  (flavor  oscillations),
$L \cdot E_\nu$  (violations  of the equivalence  principle) 
and $X \simeq L$  (flavor  changing neutral currents)
is the relevant  variable for an oscillating   transition
probability, have  been proposed as  viable  solutions 
of these  data.
Neutrino  decay    is also dependent  on the ratio
$L/E_\nu$,  but with a different functional  form.

In   this  study   we find that  flavor  oscillations
provide a significantly better  fit  to the sub--GeV and  
multi--GeV data samples  than the exotic  alternatives
we have considered,
however  with a generous  allowance for  systematic  uncertainties
the alternative  explanations  can still be   considered  as  viable.
Including the upward going  muons  in  the fit 
the alternative models are essentially ruled  out.

The upward--going muons  are  a set  of $\nu$--induced  events
corresponding to  much  larger $E_\nu$:
for  passing muons the median parent neutrino  energy is
approximately 100~GeV, with  a
significant  contribution of  neutrinos  with energy as large  as 1~TeV,
and  therefore  are in principle a powerful   handle  to  study  
the energy   dependence of the  neutrino  survival probability.
If flavor  oscillations (where $L/E_\nu$ is  the  significant
variable)  are  the cause of the suppression of 
sub--GeV  and multi--GeV muon events, 
the neutrinos  producing passing upward going   muons must also oscillate,
but  with a  smaller suppression because of their 
larger  energy;
moreover, for the range of $\Delta m^2$ suggested 
by the lower energy data, one expects  a  moderate but detectable
deformation of the zenith angle  distribution.
Both effects are  detected.

In the alternative  exotic models  we  have studied  here,
high  energy events, such as  the passing  upward--going muons,
are suppressed much more ($L\cdot E_\nu$)  than or  as much 
($X \sim L$)  as the up--going  multi--GeV  events, in contrast to  the 
experimental evidence.

Also in the  case  of neutrino decay, 
the  upward--going muon data  are very poorly fitted  by the model.
Of course   if neutrino  have  different  masses (and can  decay)  it is
natural  to expect  oscillations  in combination   with decay.
We have  not explored this  possibility,  but we can  conclude  
that  decay  cannot  be the dominant  form  of muon
neutrino disappearance.

Two  results of the measurements  of 
upward going--muons   are  critically  important  to allow
discrimination against exotic  models  and  in favour
of  usual oscillations:

\begin {itemize}
\item the     stopping/passing  (Data/Montecarlo) double  ratio  for the SK 
upward--going  muons \cite {SK-Takayama}
is $r = 0.56$,  with   a combined  statistical  and experimental 
 systematic error  of  0.07.  
The  theoretical uncertainty
in the   relative  normalization  of the two sets of data 
has  been estimated   as 8\% in \cite {upmu-LL};  more conservatively  the
SK  collaboration \cite{SK-Takayama} has  used  13\%.
Quadratically combining the   more conservative
estimate of the theoretical  uncertainty with the experimental errors,
the resulting  $\sigma_r$ is 0.1.
Therefore  the suppression for the 
high energy  passing  muons, is    weaker  than   for the lower  energy
stopping ones  at more than  four sigma  of significance,
even allowing  for  a rather large   uncertainty  in
the   theoretical  prediction.  This is  in  contrast  with
models   that predict   for the stopping/passing  double  ratio $r$
a  value of unity (flavor  changing  neutral
currents)  or larger  (violations  of the equivalence  principle).

\item  The  shape of  the through--going upward--going muons  
zenith angle  distribution shows 
indication  of  a deformation,  although   the  no--distortion 
hypothesis  (with   free normalization)  has  a  probability   close 
to 5\%.
The  deformation if  present  is a rather smooth  one, 
and the  distribution can  be  used  to rule out
models  (such  as FCNC  with smallish $\epsilon$)
that produce  deep  and marked   features in the 
neutrino distribution  (well    mapped by the  nearly collinear
muons).
\end{itemize}

The  MACRO collaboration   has  also  obtained  results
on upward--going muons \cite {MACRO-upmu},   that  indicate the presence of
an  angular  deformation
compatible  with the  presence of  flavor  oscillations
(although the oscillation fit even if  significantly better that
the   standard model  fit is  still rather  poor).  
Preliminary  results on
events  where 
upward--going  muons  are produced  in  (and exit from) the detector,
and  a   second class of  events 
that  combines stopping  upward--going  muons and 
downward--going   muons produced  in the detector
indicate  a pattern  of suppression 
that is  only  compatible  with  an oscillation  probability
that decreases  with energy \cite{MACRO-Takayama}.

Also the Kamiokande  collaboration \cite{Kamioka-up}
has  measured  passing upward--going  muons
with results in good agreement  with Super--Kamiokande,
 while  the Baksan  collaboration \cite{Baksan-up}  has   obtained  
results  not in good  agreement.
One should also note that the IMB collaboration
has in the past measured a stopping/passing ratio 
for upward--going muons in agreement with a no  oscillation
Montecarlo prediction \cite{IMB-up}  
(see \cite {LLS}  for a critical  analysis).

In conclusion, we  find  that the  present  data on atmospheric 
neutrinos  allow  to determine some
qualitative  features of the functional dependence
of the   disappearance probability  for muon neutrinos.
This probability    (smeared  by  resolution effects) 
increases  with  the pathlength $L$
producing the up/down  asymmetry  that is the strongest evidence
for  physics  beyond the standard  model.
The difference in  suppression  between the  sub(multi)--GeV
muon  events  and the higher energy through--going  muons
indicates    that the   transition probability decreases  with energy.
These results  are  in  agreement with the 
predictions of    $\nu_\mu \leftrightarrow \nu_\tau$ oscillations
and  in contrast with  several  alternative exotic models.
If flavor oscillations  are indeed  the  mechanism  for the
muon neutrinos  disappearance, 
additional  data with more  statistics  and resolution
(in $L$ and $E_\nu$)  should allow to study  in more  detail
the oscillatory structure  of the transition   probability
as  a function of  the variable $L/E_\nu$, unambiguosly
determining the  physical phenomenon.
It is natural  to expect that the 
oscillations involve  all flavors
and that electron  neutrinos  participate in the 
oscillations  (with a reduced mixing  because  of
the Chooz limit). The resulting  flavor  conversions
will have  a more  complex  dependence on the  neutrino path
and  energy $E_\nu$;  the detection   of these more  subtle
effects  could  become  the next  challenge 
for  the experimentalists.

\newpage

\begin{figure} [t]
\centerline{\psfig{figure=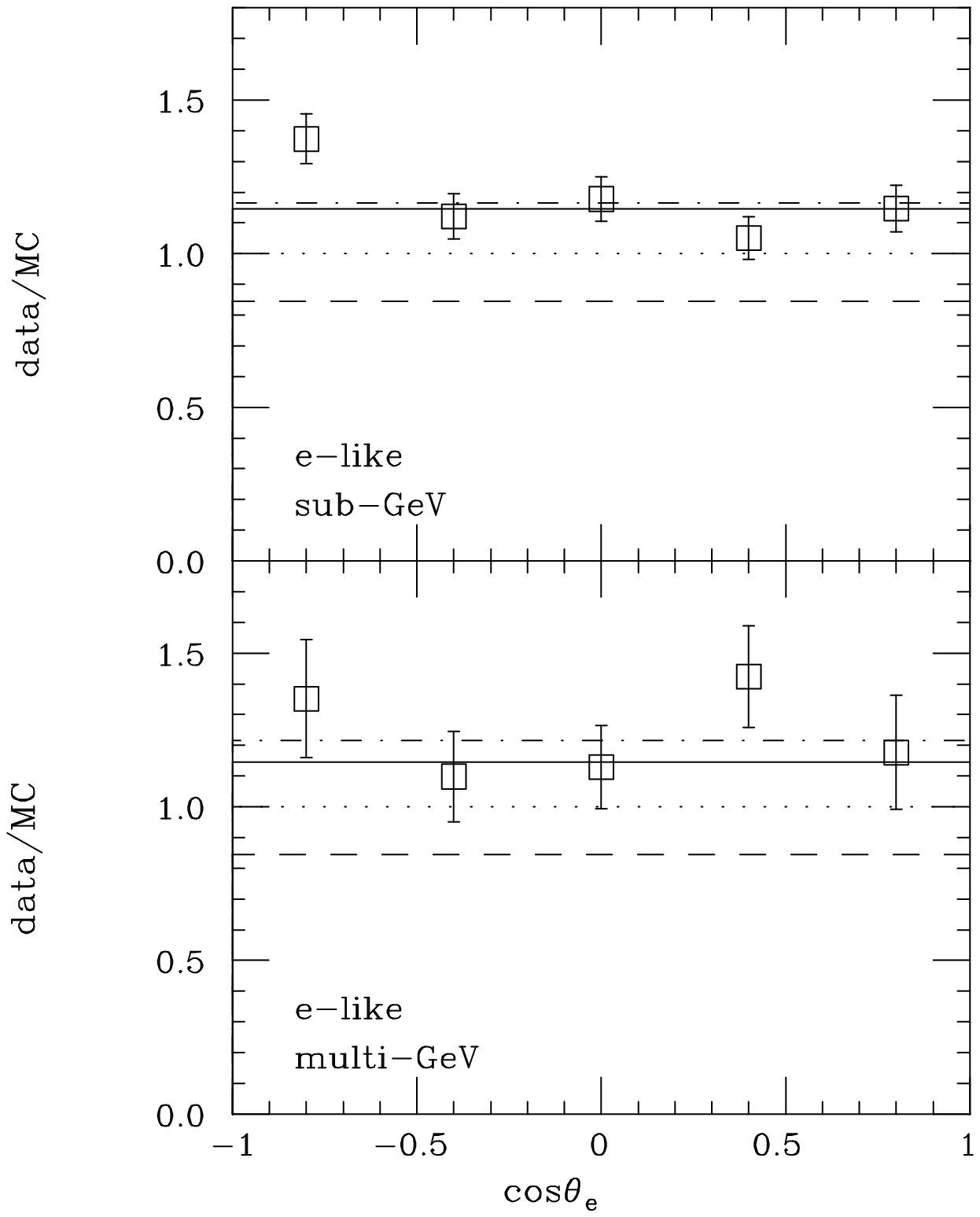,height=19cm}}
\caption {\tenrm  Ratio Data/Montecarlo for the 
$e$--like events of Super--Kamiokande.
The  dot--dashed lines are  straight line fits
(independent for each panel) to the data points;
the dashed lines  are the result of fitting all data
(included  in fig.~1, 2 and 3)  with a common  normalization
($\alpha = 0.84$); the solid  lines are the  result of
a  calculation with $\nu_\mu \leftrightarrow \nu_\tau$ oscillations
with our  best fit parameters for maximal  mixing 
($\Delta m^2 = 3.2 \times 10{-3}$~eV$^2$ and
a normalization $\alpha  = 1.15$).
}
\end{figure}

\begin{figure} [t]
\centerline{\psfig{figure=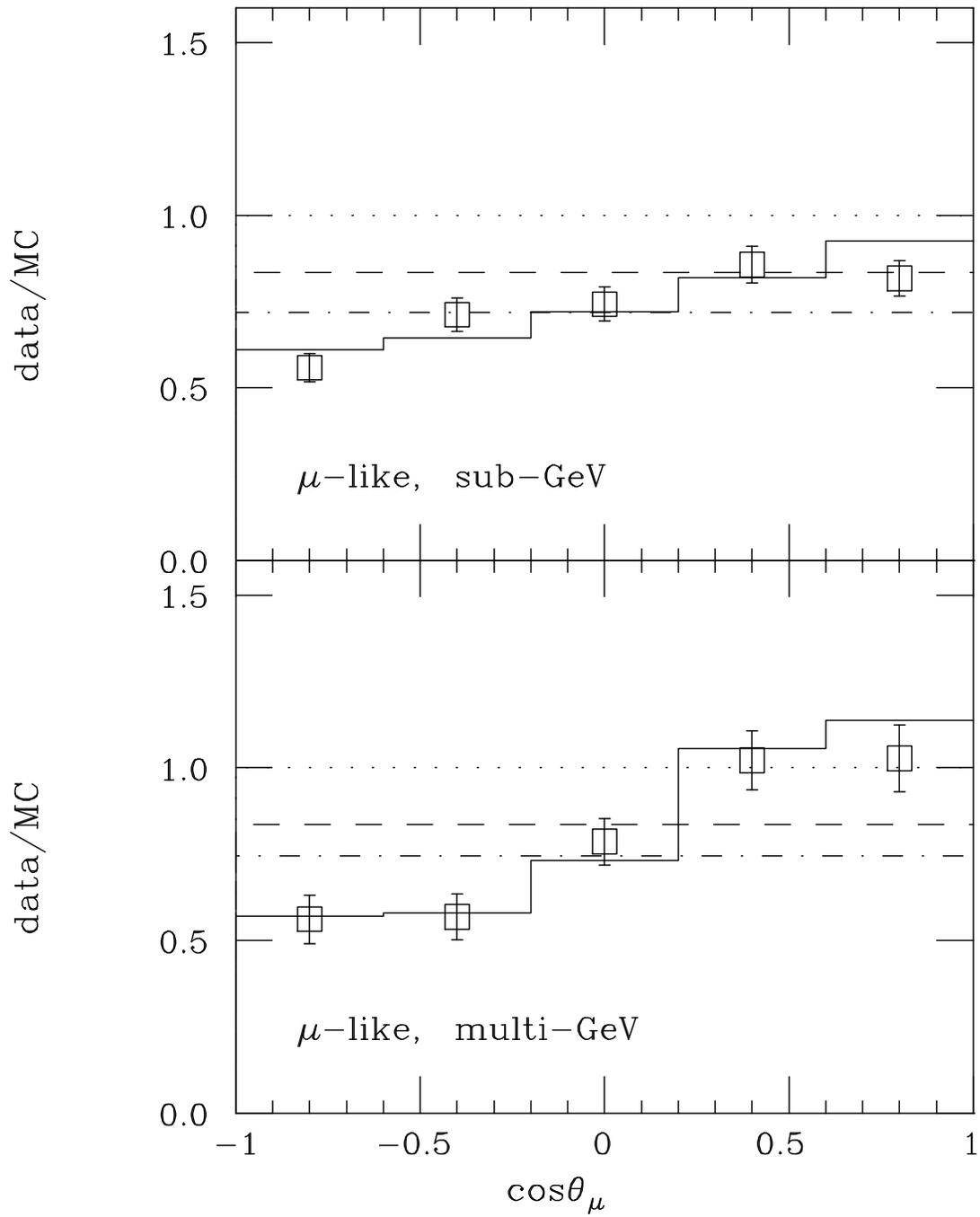,height=19cm}}
\caption {\tenrm   Ratio Data/Montecarlo for the 
$\mu$--like events of Super--Kamiokande.
See fig.~1  for a descrition of the lines.}
\end{figure}

\begin{figure} [t]
\centerline{\psfig{figure=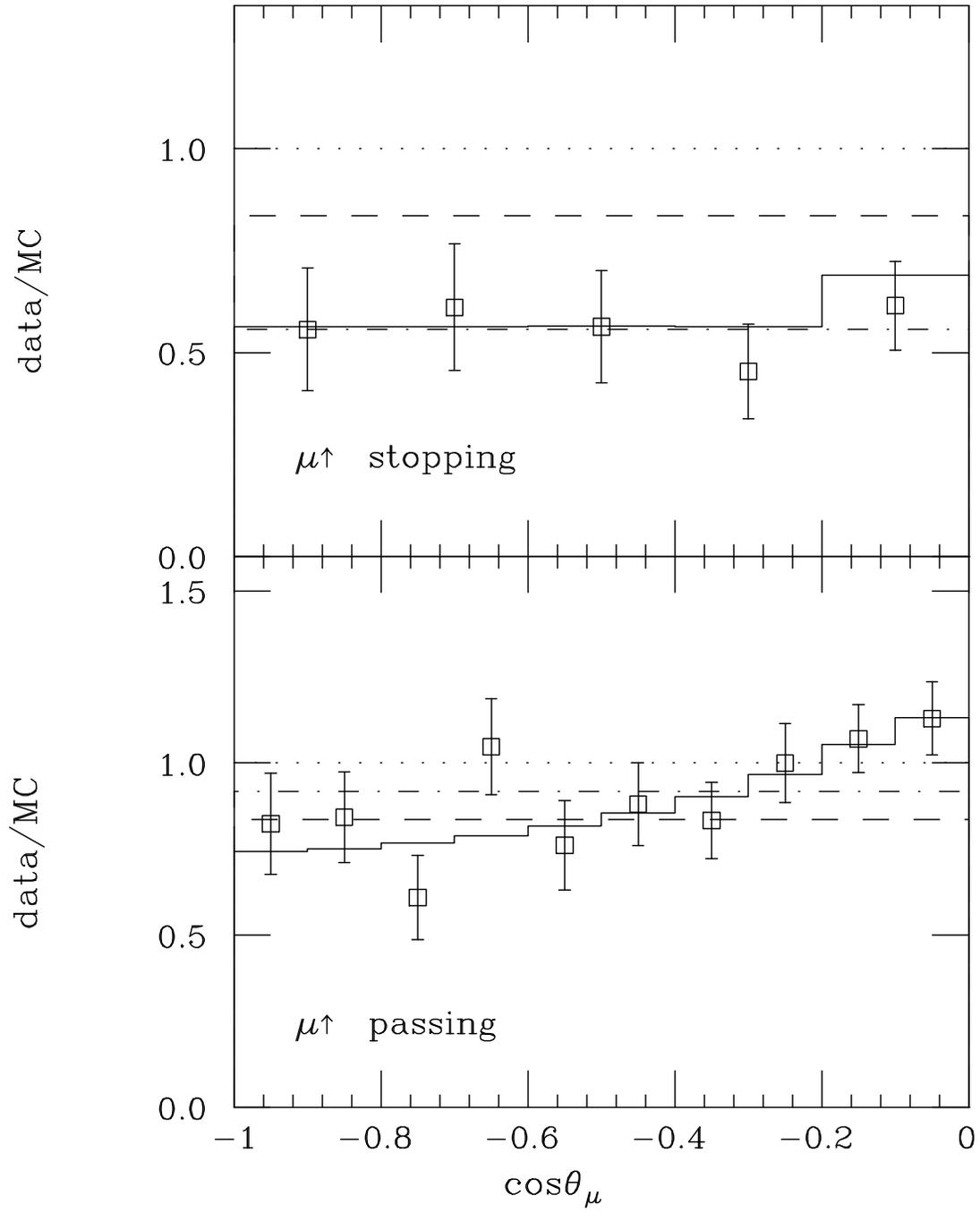,height=19cm}}
\caption {\tenrm   Ratio Data/Montecarlo for  passing
and  stopping upward--going muons in 
Super--Kamiokande. 
See fig.~1  for a descrition of the lines.}
\end{figure}

\begin{figure} [t]
\centerline{\psfig{figure=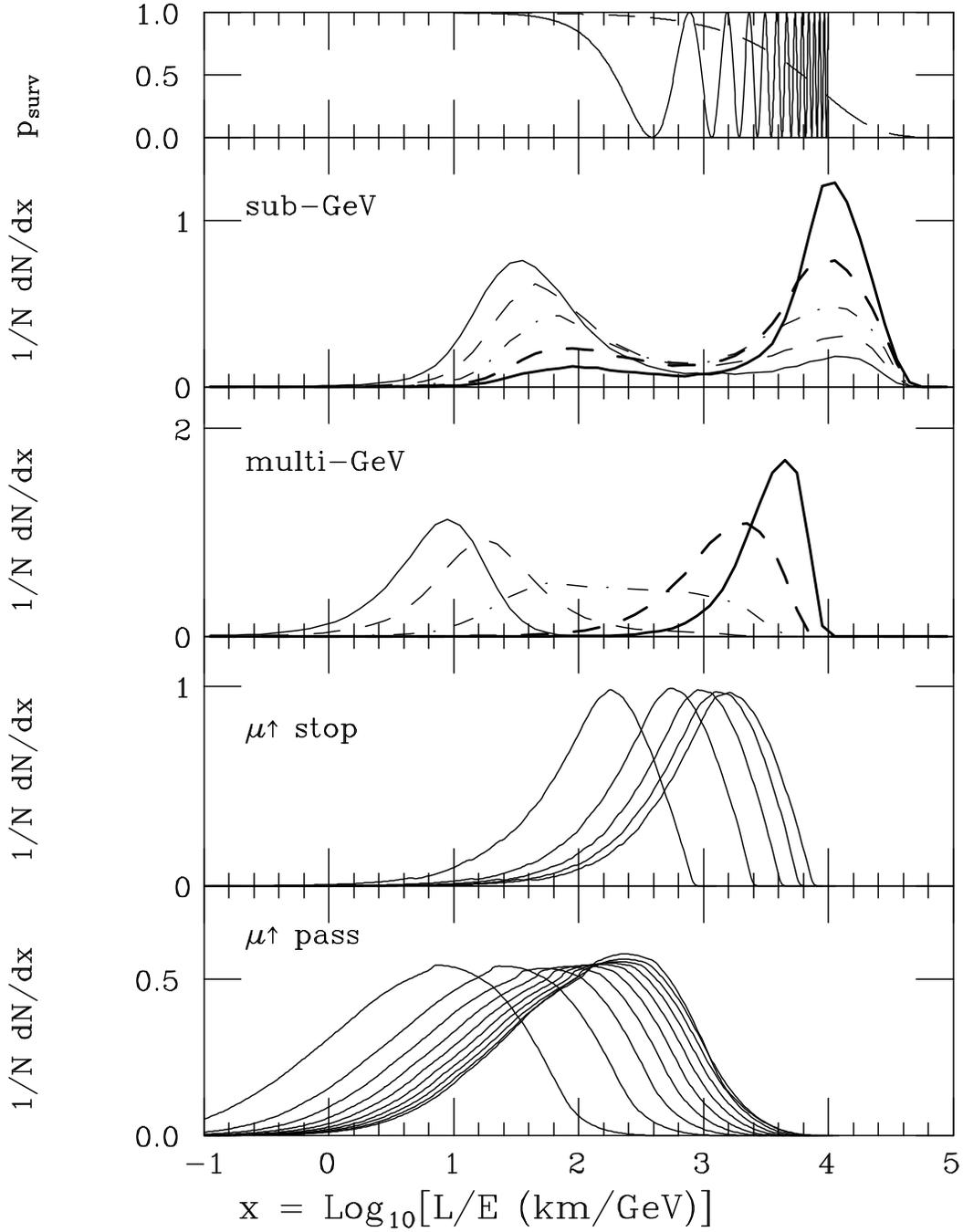,height=19cm}}
\caption {\tenrm   
Distributions in $L/E_\nu$ for the different
event  classes considered.
In the top  panel   we show the survival probability
$P(\nu_\mu \to \nu_\mu)$ for  our  best fit with flavor oscillations
to all  data (solid line) and the best fit with neutrino decay to the
sub--GeV and multi--GeV data (dashed line). }  
\end{figure}

\begin{figure} [t]
\centerline{\psfig{figure=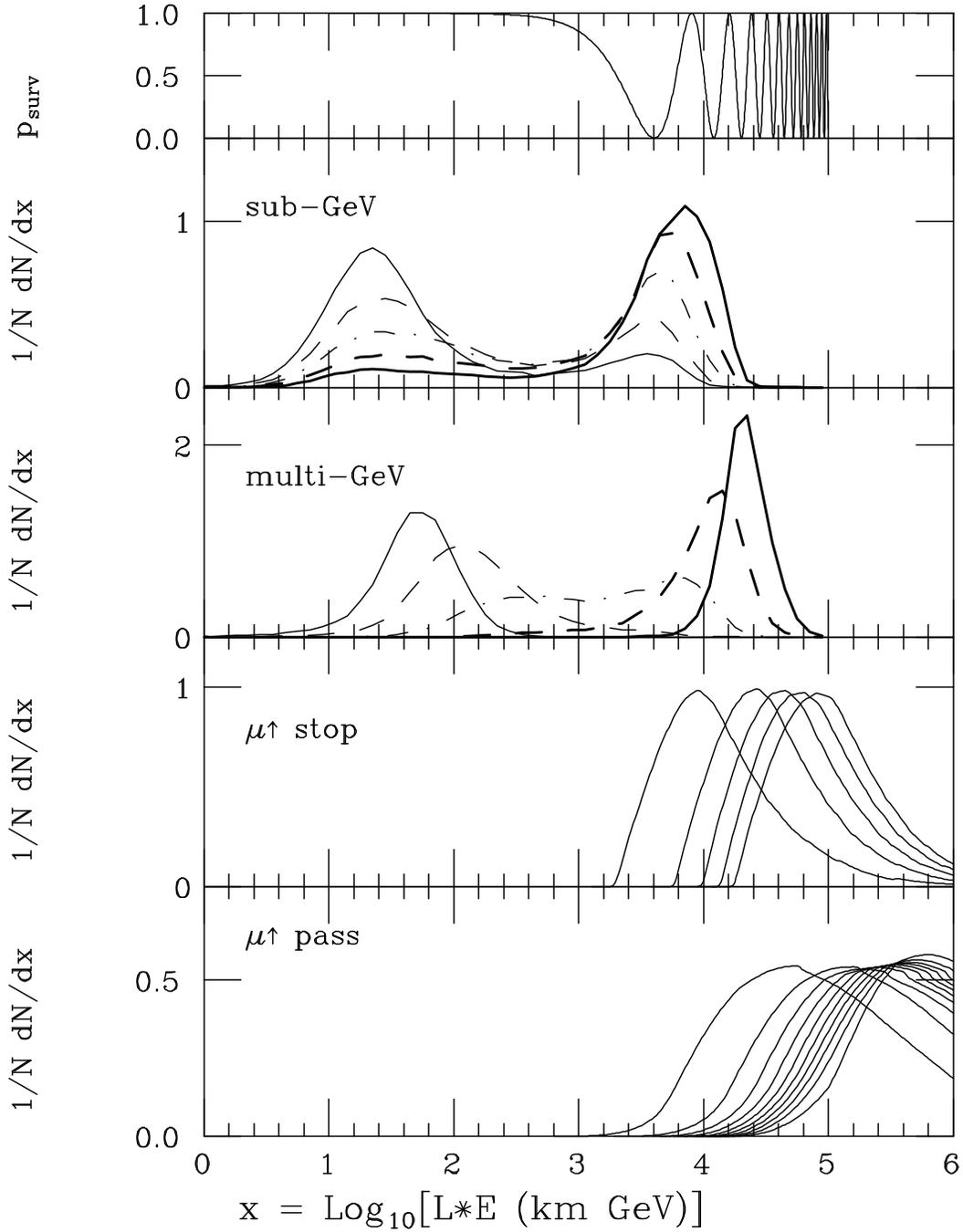,height=19cm}}
\caption {\tenrm    Distributions in $L \cdot E_\nu$ for the different
event  classes considered.
In the top  panel   we show the survival probability
$P(\nu_\mu \to \nu_\mu)$ for  our  best fit  
to the sub--GeV and multi--GeV data.}
\end{figure}

\begin{figure} [t]
\centerline{\psfig{figure=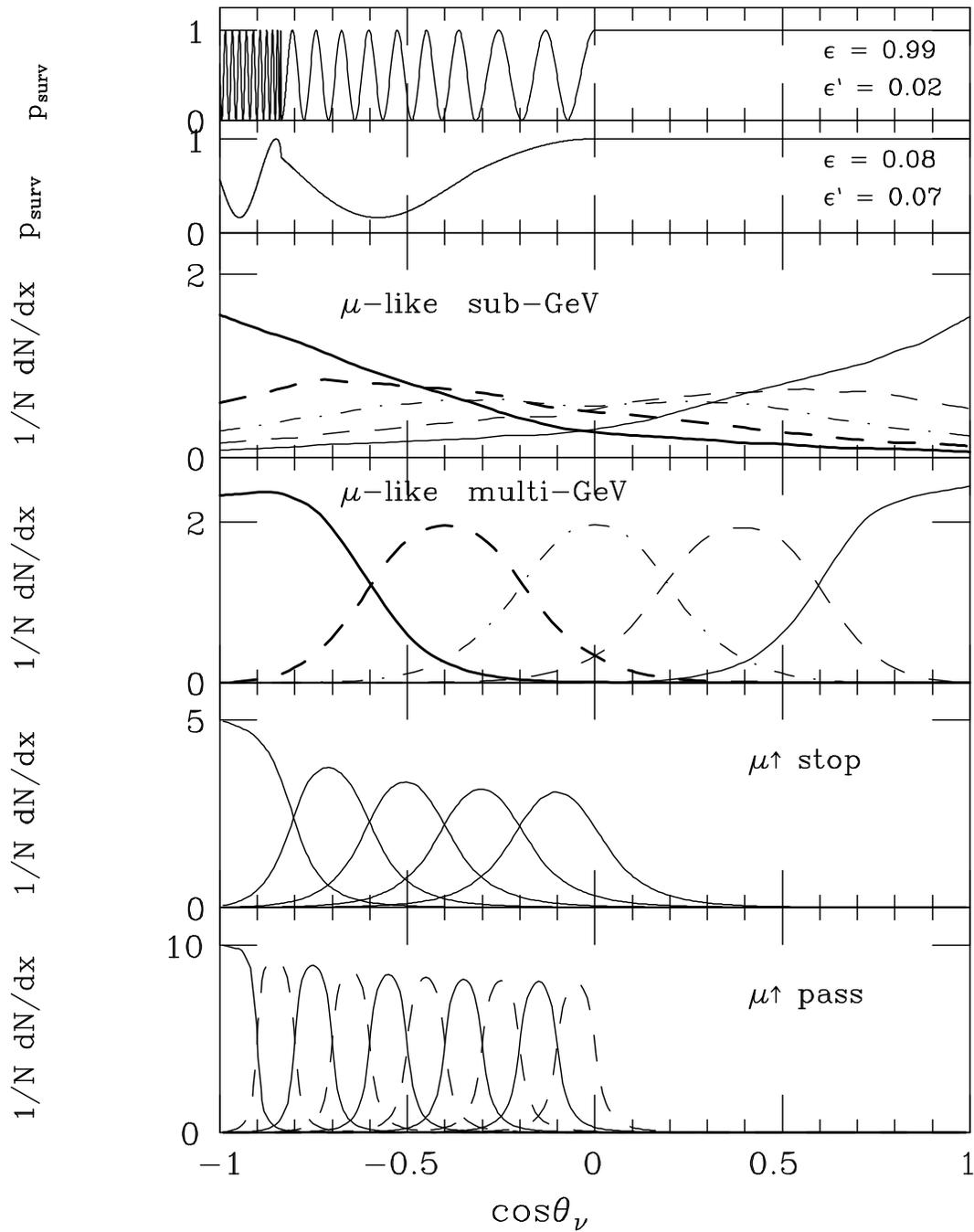,height=19cm}}
\caption {\tenrm    Distributions in $\cos \theta_\nu$ for the different
event  classes considered.
In the two top  panels   we show the survival probability
$P(\nu_\mu \to \nu_\mu)$ 
of the two  best  fit  points  as   calculated  in \protect\cite{FCNC}.
}
\end{figure}

\end{document}